\documentclass{ws-mpla}
\usepackage{epsfig,amssymb,amsmath,graphicx,color,array,arydshln,graphics,multirow}
\makeatletter
\def\hlinewd#1{%
\noalign{\ifnum0=`}\fi\hrule \@height #1 \futurelet
\reserved@a\@xhline}
\newcommand{\hthickline}{\hlinewd{1pt}}
\makeatother
\newcommand\Bpara[4]{%
\begin{picture}(0,0)%
 \setlength{\unitlength}{1pt}%
 \put(#1,#2){\rotatebox{#3}{\raisebox{0mm}[0mm][0mm]{%
            \makebox[0mm]{$\left.\rule{0mm}{#4pt}\right\}$}}}}%
\end{picture}}
\makeatletter
\def\thetable{\@arabic\c@table}
\makeatother

\begin{document}

%%%%%%%%%%%%%%%%%%%%% Publisher's Area please ignore %%%%%%%%%%%%%%
\catchline{}{}{}{}{}
%%%%%%%%%%%%%%%%%%%%%%%%%%%%%%%%%%%%%%%%%%%%%%%%%%%%%%%%%%%%%%%%%%%

\title{Inflation on a Pair of D3-brane and ${\bar {\rm D}}$3-brane\\
in Klebanov-Strassler Background}

\author{\footnotesize Yoonbai Kim, Hiroaki Nakajima, Kyungha Ryu}

\address{Department of Physics and Institute of Basic Science,
Sungkyunkwan University,\\
Suwon 440-746, Korea\\
\email{yoonbai, nakajima, eigen96@skku.edu}}

\author{\footnotesize Hang Bae Kim}

\address{Department of Physics, Hanyang University, Seoul 133-791, Korea\\
\email{hbkim@hanyang.ac.kr}}

\maketitle

%\pub{Received (Day Month Year)}{Revised (Day Month Year)}

\begin{abstract}
We explain how to obtain Klabanov-Strassler solution in the low-energy
limit of type IIB superstring theory and describe slow-roll inflation
on the system of parallely-separated D3-brane and ${\bar {\rm D}}$3-brane
in the Klebanov-Strassler background.
\end{abstract}

\ccode{PACS Nos.: 98.80.Cq, 11.25.-w}

\section{Introduction}\label{sec1}

Systematic derivation of a viable cosmological model including a natural
inflationary era in the context of superstring theory has been a challenging
problem. Particularly, in string cosmology, construction of a bridge between
string-inspired brane-world scenario based on a warped geometry and obtaining
a supergravity solution including this warped geometry has been an
attractive subject. If such supergravity solution supports nonvanishing
fluxes and is consistent with moduli stabilization, it is more intriguing.

In this note, we explain slow-roll inflation in the system of D3-brane and
anti-D3-brane in the background of Klebanov-Strassler
(KS)~\cite{Klebanov:2000hb}. In section \ref{sec2}, we briefly introduce
the massless bosonic fields in the low energy limit of type IIB superstring
theory and then summarize the KS solution involving warped geometry,
deformed conifold, constant axion-dilaton,  
and various NS-NS and R-R form fields with nonvanishing fluxes.
In section \ref{sec3}, the effective field theoretic description of
the system of D3-brane and ${\bar {\rm D}}$3-brane is given, of which action
is the sum of Dirac-Born-Infeld (DBI) type term and Wess-Zumino (WZ) type
R-R coupling. In the KS background, homogeneous time evolution of
the separated D3${\bar {\rm D}}$3 shows slow-roll inflation for a wide range
of parameter space.
We conclude this paper with a few discussions.

\section{Klebanov-Strassler Solution}\label{sec2}

In this section we briefly summarize the background geometry and fluxes
on which the system of separated D3-brane and ${\bar {\rm D}}$3-brane
lives. The specific form of (1+9)-dimensional spacetime of consideration is
from the KS solution with various fluxes and warp factor, which is obtained by
solving the supergravity equations given in the low energy limit of type
IIB superstring theory.

\subsection{IIB superstring theory and low energy limit}\label{subsec21}

In $(1+9)$ dimensions, five superstring theories are known:
\begin{center}{
\begin{tabular}{|c|c|c|c|c|}\hline
type IIB & type IIA & heterotic ${\rm E}_{8}\times {\rm E}_{8}$ &
heterotic SO(32) & type I \\
\hline
\end{tabular}
}\end{center}
We are interested in the type IIB superstring theory involving only closed
oriented strings, of which characteristic mass scale is given by the string
tension $1/\sqrt{\alpha'}$ (the corresponding string length scale is
$l_{{\rm s}}=\sqrt{2\pi\alpha'}$)
and of which mutual interaction is proportional
to the square of string coupling $g_{{\rm s}}^{2}$.

When the string tension approaches infinity $(\alpha'\rightarrow \infty)$,
all the massive modes of higher nodes decouple and
we obtain type IIB supergravity in ten dimensions as low energy
effective theory which involves only massless fields (zero modes)
and quadratic derivatives.
The bosonic sector of the closed strings is composed of six fields
completing ${\cal N}=2$ supergraviton multiplet
as summarized in Table~\ref{tab1}. %Table 1.
\begin{table}
\begin{tabular}{
!{\vrule width 1pt}>{\centering}m{17mm}!{\vrule width 1pt}>{\centering}m{17mm}
|c|>{\centering}m{15mm}|c|>{\centering}m{14mm}|>{\centering}m{16mm}!
{\vrule width 1pt}} \hthickline
& \multicolumn{6}{c!{\vrule width 1pt}}{massless bosonic fields}
\tabularnewline \hthickline
sector & \multicolumn{3}{c|}{NS-NS} & \multicolumn{3}{
c!{\vrule width 1pt}}{R-R}
\tabularnewline \hline
names & graviton & dilaton & two-form field & axion & two-form field &
{(self-dual) \\four-form\\ field}
\tabularnewline
fields & $G_{\mu \nu}$ & $\Phi$ & $B_{\mu \nu}$ & $C$ & $C_{\mu \nu}$
 & $C_{\mu \nu \rho \sigma}$ \tabularnewline
 & & & $(B_{2})$ & & $(C_{2})$ & $(C_{4})$
\tabularnewline
\rule[-3.5mm]{0mm}{8mm}{degrees} & 35 & 1 & 28 & 1 & 28 & 35
\tabularnewline
field strengths & & & $H_{\mu \nu \rho}$ \\ $(H_{3})$ & & $F_{\mu \nu \rho}$ \\
$(F_{3})$ & $F_{\mu \nu \rho \sigma \tau}$ \\ $(F_{5})$
\tabularnewline \hline
role & metric (geometry) & \multicolumn{5}{c!{\vrule width 1pt}}{matters}
\tabularnewline \hthickline
\end{tabular}
\\[1mm]
%\begin{center}{{\footnotesize
%Table 1. Massless bosonic fields in type IIB supergravity}}
%\end{center}
%\label{tab1}
\tbl{Massless bosonic fields in type IIB supergravity} 
{\begin{tabular}{m{127mm}}
\rule[0mm]{0mm}{0mm}
\end{tabular} \label{tab1}}
\end{table}

In the subsequent subsection we deal with classical equations of motion of
type IIB supergravity and obtain KS solution.

\subsection{Deformed conifold and Klebanov-Strassler background}\label{subsec22}

Superstring theories are given in (1+9) dimensions but the present Universe
we live is (1+3) dimensions. To be consistent with the observed Universe,
six spatial dimensions in a superstring theory should be unobservable and
a usual method is to assume that six spatial dimensions are compactified.
Ten-dimensional coordinates we use are given in Table~\ref{tab2}. %Table 2.
%   \setlength\dashlinedash{1pt}
%   \setlength\dashlinegap{1pt}
%
%\begin{wrapfigure}{l}{.6\textwidth}
%%  \begin{figure}
%%  \begin{center}
%%  \setlength\dashlinedash{1pt}
%%  \setlength\dashlinegap{1pt}

\begin{table}
\begin{center}{
\begin{tabular}
{|>{\centering}m{2.5mm}|>{\centering}m{2.5mm}:>{\centering}m{2.5mm}
:>{\centering}m{2.5mm}|>{\centering}m{2.5mm}:>{\centering}m{2.5mm}
:>{\centering}m{2.5mm}:>{\centering}m{2.5mm}:>{\centering}m{2.5mm}
:>{\centering}m{2.5mm}|}  
\hline
0 & 1 & 2 & 3 & 4 & 5 & 6 & 7 & 8 & 9 \rule[-2mm]{-1mm}{6mm} 
\tabularnewline \hline		   
\end{tabular} 
\begin{tabular} {m{64mm}} 
{\Bpara{30}{0}{270}{41}}{\Bpara{128}{0}{270}{61}}  
\tabularnewline 	   
\end{tabular}				
\begin{tabular} 
{m{21mm}m{39mm}} & $(~\rho~, ~~\underbrace{\theta_{1}~,~ \phi_{1}~,~ \psi~,
~ \theta_{2}~,~ \phi_{2}}_{\textstyle{\rm{compact}}}~ )$   
\tabularnewline
\rule[-2mm]{-1mm}{12mm} & 
$(~\rho~, ~~\underbrace{e_{1}~,~ e_{2}~,~ e_{3}~,~ e_{4}~,~ 
e_{5}}_{\textstyle{\rm{vielbein}}}~ )$   
\tabularnewline 
\end{tabular}		
\begin{tabular} {m{67mm}} 
%		\rule[0mm]{-2.5mm}{-0mm}
\centering $(X^{0},X^{1},X^{2},X^{3})(~\rho~, 
~~\underbrace{g_{1}~,~ g_{2}~,~ g_{3}~,~ g_{4}~,~ 
g_{5}}_{
\begin{matrix}
\rm{linear~ combination} \\%
\rm{of~ vielbein} \\%
\end{matrix}
}~ )$ 

\tabularnewline
\end{tabular}	
\begin{tabular} {m{33mm}m{9mm}m{16mm}} 
&{\Bpara{11}{4}{270}{22}}&{\Bpara{21}{4}{270}{32}}  \tabularnewline 
&\multicolumn{1}{c}{${\rm S}^2$} & \multicolumn{1}{c}{${\rm S}^3
(\mathrm{A~Cycle})$} \tabularnewline 
\end{tabular}	
\begin{tabular} {m{31mm}m{0mm}m{35mm}} 
& & {\Bpara{40}{4}{270}{51}}  \tabularnewline 
&\multicolumn{2}{c}{$T^{1,1}~\textrm{for a fixed}~ \rho \,(\neq 0)$}
\tabularnewline 
\end{tabular}
\begin{tabular} {m{23mm}m{37mm}} 
\Bpara{30}{4}{270}{41} &\Bpara{51}{4}{270}{61} \tabularnewline
\multicolumn{1}{c}{flat spacetime} & \multicolumn{1}{c}{deformed conifold} 
\tabularnewline		
\end{tabular}	
\begin{tabular} {m{64mm}} 
 {\Bpara{90}{0}{270}{100}} \tabularnewline 	
 \multicolumn{1}{c}{warped geometry}   
\end{tabular} 
%\\[1mm]
%}\end{center}	
%\begin{center}{{\footnotesize 
%Table 2. Ten-dimensional spacetime given by Klebanov-Strassler solution}}	
}\end{center}
\tbl{Ten-dimensional spacetime given by Klebanov-Strassler solution} 
{\begin{tabular}{m{127mm}}
\rule[0mm]{0mm}{0mm}
\end{tabular} \label{tab2}}

\end{table} 

%%  \end{center}
%
%%  \end{figure}
%\end{wrapfigure}
%%\begin{figure}
%%\protect\label{tab2}
%%\centerline{\includegraphics[width=60mm]{CosPfig02.eps}}
%%\caption{Ten-dimensional spacetime given by Klebanov-Strassler solution}
%%\end{figure}

\begin{figure}
\protect
\centerline{\includegraphics[width=80mm]{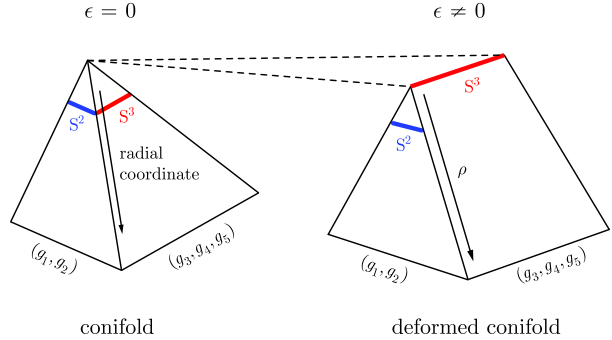}}
\caption{From conifold with $\epsilon=0$ to deformed conifold with
$\epsilon>0$.
\label{fig1}
}
\end{figure}

In the IIB superstring theory of consideration, the deformed conifold is
utilized for the construction of a tip of compact six dimensions,
of which the metric is
\begin{align}
ds^{2}_{\mathrm{def}}=\frac{1}{2}\epsilon^{\frac{4}{3}}K
\biggl[\frac{1}{3K^{3}}(d\rho^{2}+g^{2}_{5})
+\sinh^{2}\Bigl(\frac{\rho}{2}\Bigr)(g^{2}_{1}+g^{2}_{2})
+\cosh^{2}\Bigl(\frac{\rho}{2}\Bigr)(g^{2}_{3}+g^{2}_{4})\biggr],
\label{mde}
\end{align}
where the function $K$ is a function of radial coordinate $\rho$ and
decreasing,
\begin{align}
K(\rho)
&=\frac{(\sinh2\rho-2\rho)^{\frac{1}{3}}}{2^{\frac{1}{3}}\sinh\rho}
\approx
\begin{cases}
\left(\frac{2}{3}\right)^{\frac{1}{3}}\left(1-\frac{\rho^{2}}{10}
\right)+\cdots&\mbox{as $\rho\rightarrow 0$}\\
2^{\frac{1}{3}}e^{-\frac{\rho}{3}}+\cdots&\mbox{as $\rho\rightarrow
\infty$}
\end{cases},
\label{k}
\end{align}
and $g_{i}$'s $(i=1,...,5)$ stand for 
the fundamental one-forms (vielbeins) of 
the five angular coordinates.
In the metric $\epsilon$ is the deformation parameter which smooths
the singular $\mathrm{S}^{3}$ of $(g_3 ,g_4 ,g_5 )$ at the tip of conifold
as schematically shown in Fig.~\ref{fig1}.

Our (1+3)-dimensional spacetime is described by $X^{a}$'s $(a,b=0,1,2,3)$
and the metric $G_{ab}$ is assumed to be flat for the KS solution,
$G_{ab}=\eta_{ab}$.
In synthesis, the ansatz of the (1+9)-dimensional metric is
\begin{align}
ds^{2} &= {\cal H}^{-\frac{1}{2}}G_{ab}dX^{a}dX^{b} +{\cal
H}^{\frac{1}{2}}ds_{\mathrm{def}}^{2},
\label{wfm}
\end{align}
where ${\cal H}$ is a warp factor,
\begin{align}
{{\cal H}(\rho)}&=2^{\frac{2}{3}}
\epsilon^{-\frac{8}{3}}(g_{\textrm{s}}{\cal M}\alpha')^{2}I(\rho).
\label{Hrho}
\end{align}
In the warp factor ${\mathcal H}$, the constant ${\mathcal M}$ is R-R
three-form flux and $I(\rho)$ is decreasing
exponentially for large $\rho$
\begin{align}
I(\rho)&=\int_{\rho}^{\infty}dx\,\frac{x\coth x-1}{\sinh^{2}x}
(\sinh 2x-2x)^{\frac{1}{3}}\\
&\approx
\begin{cases}
I(0)-2\left(\frac{1}{6}\right)^{\frac{4}{3}}
\rho^{2}+\cdots&\mbox{as $\rho\rightarrow 0$},\quad I(0)\thickapprox0.71805
\\
3\cdot 2^{-\frac{1}{3}} \rho e^{-\frac{4}{3}\rho}+\cdots&\mbox{as
$\rho\rightarrow \infty$}
\end{cases}
.
\end{align}

\begin{figure}[h]\centering
\includegraphics[width=80mm]{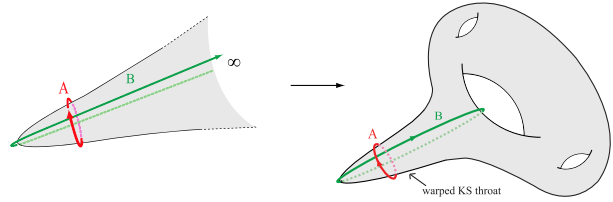}\\
\caption{From noncompact deformed conifold to compact Calabi-Yau orientifold}
\label{fig2}
\end{figure}

The KS solution involves various NS-NS and R-R form fluxes and their field
configurations are given in Table~\ref{tab3}. %Table 3.
\begin{table}
\begin{tabular}{|c|l|}\hline
field or field strength & \multicolumn{1}{c|}{solution} \\ \hline
dilaton & $~\Phi=\Phi_{0}$ \\
axion & $~C=0$ \\
\multirow{2}{35mm}[1mm]{\centering R-R three-form \\ field strength} &
$~\displaystyle{F_{3}=\frac{{\cal M}\alpha'}{2}\Bigl\{g_{5}\wedge g_{3}
\wedge g_{4}
+d\bigr[F(g_{1}\wedge g_{3}+g_{2}\wedge g_{4})\bigl]\Bigr\}}$ \\
 & $~~~~~~~\mbox{with~}\displaystyle{F(\rho)=\frac{\sinh\rho-\rho}{2\sinh\rho}}$ \\
\multirow{3}{35mm}[2mm]{\centering NS-NS two-form \\ field} &
$~\displaystyle{B_{2}=\frac{g_{\textrm{s}}{\cal M}\alpha'}{2}\left(f g_{1}
\wedge g_{2} +k g_{3}\wedge g_{4}\right)}$ \\
 & $~~~~~~~\mbox{with~}\displaystyle{f(\rho)=\frac{\rho\coth\rho-1}{2\sinh\rho}(\cosh\rho-1)}$ \\
 & $~~~~~~~\mbox{with~}\displaystyle{k(\rho)=\frac{\rho\coth\rho-1}{2\sinh\rho}(\cosh\rho+1)}$ \\
\multirow{2}{35mm}{\centering self-dual R-R five-form field strength} &
$~\tilde{F}_{5}=\mathcal{F}_{5}+\ast\mathcal{F}_{5},~~~
\mathcal{F}_{5}=B_{2}\wedge F_{3}$ \\
 & $~~~~~~~\mbox{with~}l(\rho)=f(1-F)+kF$ \\ \hline
\end{tabular}
\\[1mm]
%\begin{center}{{\footnotesize
%Table 3. NS-NS and R-R form fields and corresponding fluxes.}}
%\end{center}
%\label{tab3}

\tbl{NS-NS and R-R form fields and corresponding fluxes.} 
{\begin{tabular}{m{127mm}}
\rule[0mm]{0mm}{0mm}
\end{tabular} \label{tab3}}
\end{table}
Since the deformed conifold is not compact along the $\rho$-coordinate,
the NS-NS 3-form flux from the NS-NS 2-form field $B_{2}$ is not explicitly
given. In the phenomenological viewpoint of superstring theory, this
$\rho$-direction should also be compactified, which means that the large
$\rho$ region is chopped and is replaced by a compact geometry.
An appropriate known candidate is compact Calabi-Yau (CY) orientifold of which a
schematic shape is shown in Fig.~\ref{fig2}.
The R-R 3-form flux lives along compact $\rm A$ cycle (red line) and
the NS-NS 3-form does along compactified $\rm B$
cycle (green line) in compact CY orientifold.
This surgery is not important to describe cosmological evolution of the
early Universe which will be discussed in the subsequent section.

\section{A Pair of D-brane and ${\bar {\rm D}}$-brane in KS
background}\label{sec3}

In addition to perturbative degrees of which bosonic fields are summarized
in Table~\ref{tab1}, %Table 1,
type IIB superstring theory also involves various
branes as nonperturbative degrees. They are $p$-dimensional Dirichlet branes
(D$p$-branes), 5-dimensional Neveu-Schwarz brane (NS5-brane), and fundamental
string (F1) summarized in the following:
\begin{center}{
\begin{tabular}{|c|c|c|c|c|c|c|}\hline
(D(-1)) & D1 & D3 & D5 & D7 & NS5 & F1 \\
\hline
\end{tabular}
}\end{center}

Though each D-brane is stable and supersymmetric, a pair of D-brane and
anti-D-brane (${\bar {\rm D}}$-brane) does not possess the supersymmetry
and becomes unstable~\cite{Sen:2004nf}. Between the D$p$-brane and
${\bar {\rm D}}p$-brane, open string degrees live, of which low energy modes
are a complex tachyon field $T=\tau e^{i\chi}$ $({\bar T}=\tau e^{-i\chi})$
depicting instability, two gauge fields $A_{(n)}^{a}$ living on 
each brane, 
and two sets of transverse
coordinates $X_{(n)}^{i}$ representing the positions of D$p$-brane and
${\rm \bar{D}}p$-brane with distance
$\ell^{i}=X_{(1)}^{i}-X_{(2)}^{i}$. Dynamics of the system of D$p$-brane and
${\bar {\rm D}}p$-brane is described by effective action which consists of
DBI type term~\cite{Sen:2003tm} and WZ type R-R coupling~\cite{KKKNS},
respectively,
\begin{align}
S_{{\rm D}{\rm \bar{D}}}&=-{\cal T}_{p}\!\int\!d^{p+1}\xi \biggl[
V_{(1)}(\tau,\ell)e^{-\Phi(X_{(1)})}\sqrt{-\det A_{(1)}}
\nonumber\\
&\hspace{21mm}+
V_{(2)}(\tau,\ell)e^{-\Phi(X_{(2)})}\sqrt{-\det A_{(2)}}\,\biggr],
\label{Addb}
\\
S_{{\rm WZ}}&={\cal T}_{p}\,\int V(\tau)\,C\wedge {\rm
Str}\,e^{B_{2}+2\pi\alpha'\tilde{{\cal F}}}.
\label{DDV}
\end{align}
The D${\bar {\rm D}}$ potential in \eqref{Addb}--\eqref{DDV} is based
on the tachyon potential of an unstable D$p$-brane $V(\tau,\ell)$ as
$V_{(n)}(\tau,\ell)=V(\tau)\sqrt{\det Q_{(n)}}\,$,
and $A_{(n)}$ in the square roots are two $(1+p)\times(1+p)$
matrices as
\begin{eqnarray}\label{An}
A_{(n)ab}&=&P_{(n)ab}
\left[E_{\mu\nu}(X_{(n)})-\frac{\tau^{2}}{2\pi\alpha'\det Q_{(n)}}
E_{\mu
i}(X_{(n)})\ell^{i}\ell^{j}E_{j\nu}(X_{(n)})\right]+2\pi\alpha'F_{(n)ab}
\nonumber\\
&&+\frac{1}{\det Q_{(n)}}\biggl\{\frac{2\pi\alpha}{2} ({\overline
{D_{a}T}}D_{b}T+{\overline {D_{b}T}}D_{a}T)
\nonumber\\
&&\hspace{15mm}
+\frac{i}{2}\left[E_{ai}(X_{(n)})+\partial_{a}X_{(n)j}E_{ji}(X_{(n)})\right]
\ell^{i}(T{\overline {D_{b}T}}-{\bar T}D_{b}T)
\nonumber\\
&&\hspace{15mm} +\frac{i}{2}(T{\overline {D_{a}T}}-{\bar
T}D_{a}T)\ell^{i}
\left[E_{ib}(X_{(n)})-E_{ij}(X_{(n)})\partial_{b}X_{(n)j}\right]
\biggr\}.
\end{eqnarray}
In the previous expression, $P_{(n)}^{ab}[\cdots]$ means pull-back
of the closed string fields on the $n$-th brane,
$E_{\mu\nu}=G_{\mu\nu}+B_{\mu\nu}$ raises and lowers the indices in
the action, and
\begin{eqnarray}\label{det}
\det
Q_{(n)}=1+\frac{\tau^{2}}{2\pi\alpha}\ell^{i}\ell^{j}G_{ij}(X_{(n)}).
\end{eqnarray}
The field strength tensor of a U(1) gauge field on the $n$-th brane
is $F_{(n)}^{ab}=\partial^{a}A_{(n)}^{b}-\partial^{b}A_{(n)}^{a}$
and the covariant derivative of complex tachyon field is
$D^{a}T=\partial^{a}T-i(A_{(1)}^{a}-A_{(2)}^{a})T$.
In \eqref{DDV}, Str denotes supertrace and
\begin{equation}\label{TF}
\tilde{{\cal F}} = \left(
\begin{array}{cc}
F_{(1)}-{\rm i}_{d\Phi_{1}}\,\,\,\,
& i^{3/2}\left[DT+iT({\rm i}_{\Phi_{1}}-{\rm i}_{\Phi_{2}})\right] \\
i^{3/2}\left[{ \overline{DT}}-i{\bar T}({\rm i}_{\Phi_{1}}-{\rm
i}_{\Phi_{2}})\right]\,\,\,\, & F_{(2)}-{\rm i}_{d\Phi_{2}}
\end{array}
\right),
\end{equation}
where ${\rm i}_{\Phi_{n}}$ denotes the interior product by $\Phi_{n}$ regarded
as a vector in the transverse space.

\begin{figure}\centering
\scalebox{0.35}[0.35]{\includegraphics{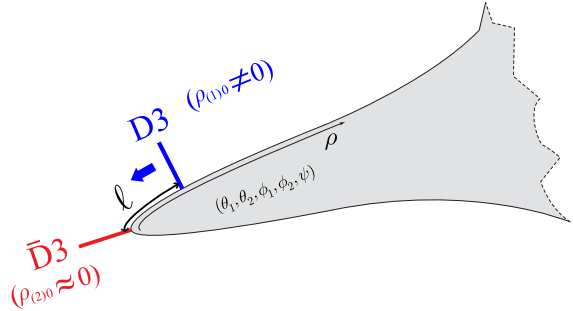}}\\
\caption{D3 at $\rho=\rho_{(1)0}\gtrsim l_s$ (blue color) and ${\bar
{\rm D}}$3 at $\rho=\rho_{(2)0}\approx0$ (red color) in a warped
throat of the deformed conifold.} \label{fig3}
\end{figure}

Now we introduce a D3-brane and an ${\bar {\rm D}}$3-brane in the KS
background.
If we consider the toal action $S_{{\rm D}{\bar {\rm D}}}+S_{{\rm WZ}}$,
the potential terms of the D3-brane
inversely proportional to the warp factor, ${\cal H}^{-1}(\rho_{(1)})$,
do not appear due to
cancellation between the contribution from the DBI type action
\eqref{Addb} (or the NS-NS coupling) and that from the WZ type action
\eqref{DDV} (or the R-R
coupling), while the contributions are added up for the $\bar {\rm
D}$3-brane, proportional to $2{\cal H}^{-1}(\rho_{(2)})$. It means that
the D3-brane experiences no net force from the background, but the
$\bar{\textrm{D}}3$-brane does the attractive force from the
background. Resultantly, as a natural initial configuration, the D3-brane
is located at a position of the warped throat, while the $\bar
{\rm D}$3-brane is located at the tip of the
deformed conifold as shown in Fig.~\ref{fig3}.
Here we introduce the distance $\ell$ between the D3-brane and the
anti-D3-brane as
\begin{equation}
\ell^{i}=
\begin{cases}
\rho_{(1)}-\rho_{(2)}\equiv{\ell}\mbox{\quad for $i=\rho$}\\
0\mbox{\qquad\qquad\qquad\, otherwise}
\end{cases}
.
\end{equation}

In order to study cosmological implication of the D${\bar {\rm D}}$
system, of which main topic is realization of inflationary era, we
choose the static gauge $\xi^{a}=X^{a}$, assume absence of nontrivial
gauge fields, $A_{(n)}^{a}=0$, and consider homogeneous open string fields
\begin{equation}
\tau=\tau(t),\quad \chi=\chi(t),\quad \ell=\ell(t).
\end{equation}
In the world-volume of D${\bar {\rm D}}$ system
we assume the flat Robertson-Walker metric
\begin{align}\label{RW}
ds^{2}=-dt^{2}+a^{2}(t)\left[\frac{dr^{2}}{1-{\rm k}
r^2}+r^2(d\theta^2+\sin^2 \theta d \phi^2)\right],
\qquad ({\rm k}=0),
\end{align}
instead of the (1+3)-dimensional Minkowski spacetime in \eqref{wfm}.
In the 6-dimensional space of deformed conifold,
we also read and substitute the string coupling $g_{{\rm s}}=e^{\Phi_{0}}$,
the trivial axion field $C=0$,
nonvanishing NS-NS 2-form field $B_{2}$, R-R 2-form field $C_{2}$, and
R-R 4-form field $C_{4}$ from the self-dual R-R 5-form field ${\tilde F}_{5}$,
summarized in Table~\ref{tab3}.

Dynamics of the closed string degrees in the weak coupling limit is
of order $1/g_{{\rm s}}^{2}$ but that of open string degrees is of
order ${\cal T}_{p}= g_{{\rm s}}^{-1}(2\pi)^{-p}(\alpha')^{-\frac{p+1}{2}}$.
Therefore, the interaction between the D$p$-brane and ${\bar {\rm D}}p$-brane
from the closed string degrees which is a
$1/g_{{\rm s}}$ correction should be
taken into account. When the transverse distance is large enough
$(\ell > \ell_{{\rm c}})$, the 1-loop correction~\cite{Banks:1995ch}
provides ${\cal O}(1/\ell^{4})$ order corrections from gravitation
and R-R coupling of which magnitudes and signatures are exactly the
same. As the distance between D and ${\bar {\rm D}}$ reaches a
critical distance, the 1-loop amplitude diverges. A natural
assumption for the coincident D${\bar {\rm D}}$ $(\ell =0)$ is to
introduce a finite binding energy per unit area ${\cal E}_{{\rm
b}}$. Interpolation of both limits suggests the
following correction to the tachyon potential in the last square bracket
\begin{align}
V_{(1)}(\tau,\ell)=V_{(2)}(\tau,\ell)=V(\tau,\ell)=
\frac{1}{\cosh(\sqrt{\pi}\,\tau)}
\sqrt{1+\frac{\tau^{2}\ell^{2}}{2\pi\alpha'}}
\left[1-\frac{\mathcal{E}_{\mathrm{b}}/2\mathcal{T}_{3}}
{1+(\ell/\ell_{\mathrm{c}})^{4}}\right],
\label{cop}
\end{align}
reflecting the gravitational and R-R attractions between the D-brane
and ${\bar {\rm D}}$-brane. In (\ref{cop}), $\ell_{{\rm
c}}=\left(2\kappa_{10}^2{\cal T}_3^2/{\cal E}_{b}\right)^{1/4}$,
where $\kappa_{10}^2$ is ten-dimensional gravitational constant.

In Fig.~\ref{fig3}, the buoyant D3-brane starts to move to the sunken
$\bar{\textrm{D}}$3-brane by these atrractive forces, however they are
weak enough due to the nontrivial warp factor.
We assume that $\rho_{(2)}$ is always sufficiently small.
Furthermore, to perform the numerical analysis, we set the position
of anti D3-brane $\rho_{(2)}$ to be fixed at the tip of warped
throat in the KS background, which naturally gives
$\dot{\rho}_{(2)}=0$. Note that we consider
only the motion along the radial coordinate $\rho$
and omit the dynamics of the angular variables
$(\theta_{1},\theta_{2},\phi_{1},\phi_{2},\psi)$ in
the deformed conifold for simplicity, which may not lose generality.
In addition, we do not consider dynamics of the tachyon phase field
$\chi(t)=0$.

To perform the numerical
analysis for the slow-roll inflation in the D${\bar {\rm D}}$ system, we
introduce the dimensionless quantities
\begin{equation}
\frac{t}{l_{\mathrm{s}}},\quad a,\quad \tau,\quad
\frac{\rho_{(1)}}{l_{\mathrm{s}}},\quad {\tilde {\cal T}}_{3}=
\frac{\mathcal{T}_{3}l_{\mathrm{s}}^{2}}{M_\mathrm{P}^{2}},\quad
e_{\mathrm{b}}=\frac{\mathcal{E}_{\mathrm{b}}}{\mathcal{T}_{3}},\quad
\frac{\ell_{\mathrm{c}}}{l_{\mathrm{s}}},\quad {\cal M}, \quad
\epsilon=\exp\left(-\frac{\pi {\cal K}}{g_{\mathrm{s}}{\cal
M}}\right),
\end{equation}
where $M_{{\rm P}}$ is Planck mass and
the deformation parameter
$\epsilon$ is given by the R-R 3-form flux $\cal M$ and NS-NS 3-form
flux $\mathcal{K}$.
An appropriate set of initial conditions is
\begin{eqnarray}
\tau(0)=\tau_{0},\quad \ell(0)=\ell_{0},\quad {\dot \tau}(0)={\dot
\tau}_{0},\quad {\dot \ell}(0)={\dot \ell}_{0}, \label{icd}
\end{eqnarray}
since $a(0)$ can always be fixed to be unity for ${\rm k}=0$ from
the Einstein equations. Actual numerical
analysis will be performed under a natural condition of no initial
time derivatives, ${\dot \tau}_{0}={\dot \ell}_{0}=0$. In synthesis,
the expansion rate represented by the value of e-folding is studied
in the space of four dimensionless parameters,
$({\tilde {\cal T}}_{3}, e_{{\rm b}},\tau_{0},\ell_{0})$.

\begin{figure}\centering
\scalebox{0.44}[0.44]{\includegraphics{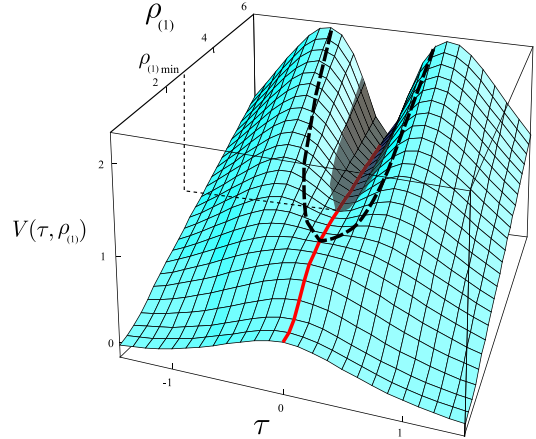}}\\
\caption{The tachyon potential with
$\mathcal{E}_{\mathrm{b}}=0.5$, ${\cal T}_{3}=1$,
$\ell_{\textrm{c}}=1$, and $\rho_{(1)\rm min}=3.18$.}
\label{fig4}
\end{figure}

As shown in Figure~\ref{fig4}, the shaded area
by thick-grey color stands for the region of sufficient slow-roll
inflation over 60 e-folding in the KS background.
For comparison, the area bounded by the dashed line
represents the region of 60 e-folding condition in the flat
background. As the tension ${\cal T}_{3}$ increases, the initial
distance $\ell_{0}$ decreases and the initial tachyon amplitude
$\tau_{0}$ increases.

\section{Discussion}\label{sec4}

The KS solution has constant dilaton and axion configuration, and it
is proven to be a dynamically-favorable configuration~\cite{Giddings:2001yu}
and other moduli including the volume moduli can also be
stabilized by adding D7-branes and $\overline{\rm D3}$-branes~\cite{Kachru:2003aw}.
Though we obtained a slow-roll inflation model based on D${\bar {\rm D}}$
system in the IIB superstring thoery~\cite{Kachru:2003sx}, it is suffered by
huge supergravity corrections, and thus
its present form does not generate a natural inflationary
era~\cite{Baumann:2007np}.

When the fluxes were also generated on the D3-brane and are left in the present
Universe, it may also jeopardize the cosmological model.
As far as the dilaton moduli is fixed, they can
sufficiently be diluted through the cosmological expansion~\cite{Chun:2005ee}.
Since this result is obtained without R-R form fluxes, an intriguing
research direction may be inclusion of fluxes both in D-brane and compactified
directions.

\section*{Acknowledgment}
The authors would like to appreciate the helpful discussions with
Akira Ishida, Taekyung Kim, and Jin Hyang Song. This work was
supported by the Korea Research Foundation Grant funded by the
Korean Government (KRF-2008-313-C00170) (Y.K.),
Astrophysical Research Center for the Structure and Evolution of the
Cosmos (ARCSEC) (K.R., H.N.), the research fund of Hanyang University
(HY-2006-S) (H.B.K).

\bibliographystyle{aipproc}   % if natbib is available

\end{document}